# Single ice crystal growth with controlled orientation during directional freezing

*Tongxin Zhang, Lilin Wang\*, Zhijun Wang\*, Junjie Li, Jincheng Wang*

*State Key Laboratory of Solidification Processing, Northwestern Polytechnical University, Xi'an 710072, China*

**Abstract:** Ice growth has attracted great attention for its capability of fabricating hierarchically porous microstructure. However, the formation of tilted lamellar microstructure during freezing needs to be reconsidered due to the limited control of ice orientation with respect to thermal gradient during in-situ observations, which can greatly enrich our insight into architectural control of porous biomaterials. This paper provides an in-situ study of solid/liquid interface morphology evolution of directionally solidified single crystal ice with its C-axis (optical axis) perpendicular to directions of both thermal gradient and incident light in poly (vinyl alcohol, PVA) solutions. Misty morphology and V-shaped lamellar morphology were clearly observed in-situ for the first time. Quantitative characterizations on lamellar spacing, tilt angle and tip undercooling of lamellar ice platelets provide a clearer insight into the inherent ice growth habit in polymeric aqueous systems and are suggested exert significant impact on future design and optimization in porous biomaterials.

## I. Introduction

Ice templating is a novel technique for fabricating hierarchically porous materials. It is also famous for mimicking nacre-like composite materials with high yield strength and fracture toughness by infiltrating alloys [1-2] and plastics [3] into the ice-templated ceramic scaffold. Very recently, ice templating has been extended to more diverse fields such as solid oxide cell [4], bio-inspired polymeric woods [5] and hierarchically structured aerogel [6]. In ice templating, the tailored hierarchically porous microstructure is the most compelling feature, and the tunability of porous microstructure through water freezing has been one of the fundamental issues in ice templating. Despite many previous efforts, the microstructure features need to be further in-situ investigated in terms of its relation with thermal gradient on the single ice crystal level to uncover the pattern formation mechanism in ice templating.

---
\* Corresponding author. wlilin@nwpu.edu.cn
\* Corresponding author. zhjwang@nwpu.edu.cn

The in-situ observation is the most direct and helpful way to reveal the pattern formation in ice templating. In recent in-situ observations of the S/L interface morphology evolution of ice with well-designed experimental set-up [7] [8-10], the diversity of the microstructure is frustrating. The crystalline orientation of ice is the key factor to illustrate the diverse microstructure formation [11-12]. In directional solidification, the S/L interface microstructure depends on the crystallographic orientation with respect to the direction of thermal gradient. However, in all previous in-situ experiments, the ice crystal orientation was not well manipulated with respect to the thermal gradient, which has become a big barrier to further investigations on lamellar microstructure evolution in ice templating. Since ice platelets originate from the instability of edge plane of ice, the morphology evolution of randomly oriented ice in a sample cannot provide quantitative information. In order to answer the long debating question on tilted growth of directionally solidified ice in aqueous polymer solutions and colloidal suspensions, we need to in-situ observe the formation of lamellar microstructure of ice with its preferred orientation parallel to the thermal gradient. The directional growth with their C-axes perpendicular to directions of both thermal gradient and incident light is a precondition for answering the long debating question on tilted growth of directionally solidified ice. In fact, it is a great challenge to control the ice crystal orientation during in-situ observations in ice templating.

Apart from the crystal orientation effect, the morphology instability greatly depends on the solute field via presence of various additives in ice templating. In solidification theory, the constitutional undercooling is the most direct inducement for an initially planar S/L interface to become instable. It was believed [9] that linear stability analysis could provide quantitative explanation of the characteristic scale of lamellar spacing via specific instability wavelength of polymer solution. Then, the physics of S/L interface instability behind polymer solutions with in-situ quantitative measurement would be very intriguing compared to aqueous solutions of small molecules, where the interface undercooling is an important parameter. However, the quantitative measurement of interface undercooling is still absent. Very recently, a new method has been proposed to in-situ measure the interface undercooling during directional freezing of colloidal suspensions [13]. The measurement distinguished the difference between particulate undercooling and constitutional undercooling [14] and further revealed the instability mode of S/L interface in colloidal suspensions [15]. Accordingly, the interface undercooling in directional freezing of polymer solutions is an important indicator to enriching the understanding of the S/L interface microstructure evolution in ice templating.

It is necessary to obtain dynamic evolution information in terms of S/L interface instability, interface undercooling and lamellar spacing adjustment in microstructure evolution. Although some relevant studies provided useful information on freezing of polymer solutions, in-situ microstructure evolution information in connection with its tip undercooling, primary lamellar spacing and other thermodynamic properties is not clear yet. Two major difficulties prevent further advances in this problem. One is the unique properties of polymer solutions [16] which arise from the nonlinear dependence of mass transport [17-20], interface attachment processes [21-24] and freezing point depression[25-26] on polymer concentration and its other chemical properties. The other one stems from the limited experimental techniques to control the crystal orientation of ice sample with respect to thermal gradient with in-situ observation since lamellar ice platelets originate from the instability of edge plane of ice.

Quantitative information is strongly required in further decoding the microstructure evolution in ice templating. To understand how lamellar microstructure is developed, it is of significant interests to obtain new in-situ experimental observations with varying growth conditions like growth velocity, solute concentration and chemical property of polymer solute for fixed ice crystal orientation. In this paper, we in-situ investigated the S/L interface microstructure evolution of unidirectionally solidified PVA solutions with well-defined ice crystal orientation and revealed the effect of growth velocity, solute concentration and degree of hydrolysis of PVA on it.

## II. Experimental details

The experiments were performed in a unidirectional freezing manner with each single ice crystal of the same crystal orientation grown in a rectangular capillary tube. **Figure 1** show the control principle of crystal orientation with birefringence of ice and the schematic diagram of unidirectional freezing with measurement of tip undercooling. It was reported by Brewster [27-28] that ice is a "positive uniaxial doubly-refracting" material due to its hexagonal crystal symmetry, and any incident light whose direction is not parallel to the C-axis of ice crystal can be "resolved into ordinary and extraordinary components" through ice crystal [29]. It has been shown in *Physics of ice* that [29] "the birefringence of ice is extremely useful for determining the orientations of the C-axis in crystals and observing the grain structure in thin sections of poly-crystalline ice".

Based on crystal optics, the specific orientation relation of an ice crystal with respect to the laboratory frame *A-P-L* (*A-A* is the direction of analyzer, *P-P* is the direction of polarizer and *L* is the direction of incident polarized light) is directly linked with the dimensionless intensity $I_\perp/I_0$ of incident polarized light, which is determined by both α and β with three relations (Eq.1- Eq.3), as illustrated in **Figure 1(a)**.

$$\text{for } 0 \leq \alpha \leq 90° : \Delta N_\alpha = \sqrt{\dfrac{1}{\dfrac{\cos^2 \alpha}{N_0^{\ 2}} + \dfrac{\sin^2 \alpha}{N_e^{\ 2}}}} - N_0 \qquad \text{(Eq. 1)}$$

$$R = \Delta N_\alpha \cdot d \qquad \text{(Eq. 2)}$$

$$I_\perp / I_0 = \sin^2(\frac{\pi R}{\lambda}) \cdot \sin^2 2\beta \qquad \text{(Eq. 3)}$$

where α defined as "tilt angle of optical axis" is an acute angle of C-axis of ice crystal tilting from the incident light direction **L**; β defined as "extinction angle" is an angle between the projection direction of C-axis in **A-P** plane and the direction **P-P**; $N_0$ and $N_e$ are refractive indexes for ordinary and extraordinary light through ice; $\Delta N_\alpha$ is the birefractive index of ice with a corresponding $\alpha$; $R$ is the optical path difference of ordinary and extraordinary light due to the birefringence of ice; $d$ is the thickness of ice crystal being transmitted; $I_0$ is the incident intensity of polarized light and $I_\perp$ is the transmitted intensity of polarized light; $\lambda$ is the wavelength of polarized light.

The $\Delta N_\alpha$-α curve from Eq.1 is plotted in **Figure 1(b),** where the birefractive index $\Delta N_\alpha$ monotonically increases with tilt angle of optical axis α (0≤α≤90°) to a maximum value with a corresponding position labeled as "*M*". According to Eq.3, the dimensionless intensity $I_\perp / I_0$ which corresponds to the length of line "*OF*" against extinction angle β on polar coordinate system exhibits a quartic symmetry, as plotted in **Figure 1(c)**. When β = 0°, 90°, 180° or 270°, $I_\perp / I_0 = 0$, extinction will occur in which ice sample appears dark and such direction is called "extinction direction". α and β can be manipulated by changing the position of the specimen box fixed to the frame **X-Y-Z** where the ice crystal grows with respect to the frame **A-P-L**. By manipulating the two parameters α and β of the ice crystal to specific values based on the orientation relation between the frame **X-Y-Z** and the laboratory frame **A-P-L**, the dark "extinction position" (noted as "$EI$") can be determined and one can finally obtain an ice crystal whose C-axis is perpendicular to both the thermal gradient and incident light for the following directional freezing experiments. A step-by-step methodology based on crystal optics is graphically illustrated in **Figure 1(d-g)** [30]. A single ice crystal was guaranteed by uniformly dark image in every step when rotating the specimen box under crossed polarizers because grain boundaries of poly-crystalline ice can be recognized if it does not appear uniformly dark in the extinction position [29].

**Figure 1(h)** shows the schematic diagram of horizontal directional freezing stage and measurement of tip undercooling $\Delta T_{tip}$ by differential visualization (DV) method [13-14] (more details on determination of the tip undercooling and its error by DV method are given in **Auxiliary Supplementary Materials 1**). In each capillary tube prior to in-situ directional freezing, the C-axis of single ice crystal was manipulated to be perpendicular to directions of both the thermal gradient and incident light. We select this direction of C-axis for our experiments since the directional growth with their C-axes perpendicular to directions of both thermal gradient and incident light is a precondition for answering the long debating question on tilted growth of directionally solidified ice. The imposed temperature gradient for directional growth was $G = 5.0 \pm 0.8$ K/mm. The microstructure evolution of S/L interface was recorded by a CCD camera. In addition, the ice crystal orientation was simultaneously detected through a pair of polarizers to guarantee that the crystal orientation remained unchanged during freezing of PVA solutions. Tip undercoolings under different pulling velocities for all samples were precisely obtained by DV method [13-14].

Poly (vinyl alcohol, PVA) is usually added to solutions in small amounts as a binder [31-32] or large amounts as a scaffold [33-34] to form aligned porous materials. The lamellar or fishbone-like microstructure mainly depends on the freezing rate [33], temperature gradient [35] and polymer concentration [31-32]. PVA solution belongs to a complex system with peculiar properties which distinctively differs from small-molecule aqueous solutions. For example, its widely investigated properties of interface adsorption [36], phase separation [37] and sol-gel transition [38] rely heavily on many factors like its molecular weight, mass fraction and other chemical properties, which are suggested to inherently affect the freezing process. However, Previous reports [39-42] mainly focused on the effect of polymer concentration and/or polymer size on microstructures of ice templating samples. Among these factors, its degree of hydrolysis, as an important chemical property, has been paid little attention in most ice templating experiments. In view of these considerations, two types of PVA solutions (PVA103 (96.8%-97.6% hydrolyzed) and PVA203 (86.5%-89% hydrolyzed) were directionally solidified in capillary tubes with four different concentrations (0.25 wt.%, 1 wt.%, 5 wt.% and 10 wt.%) under increasing pulling velocity approximately ranging from 3.5 um/s to 57 um/s, where each ice crystal in their capillary tubes shared the same orientation relation with respect to the frame **A-P-L** (For example, see a speed-up version of microstructure evolution of 5 wt. % PVA103 solutions with simultaneous measurement of tip undercooling by DV method under increasing growth velocities in **Auxiliary Supplementary Materials 2**). PVA solutions were prepared at 80℃ by dissolving PVA powder in ultrapure water (provided by deionizer when the resistance of water came to 18.25 MΩ). Rectangular glass capillary tube with an inner space dimension of $0.05 \times 1$ mm$^2$ (VitroCom brand) was adopted as sample cell for unidirectional freezing of PVA solutions. Prior to ice growth experiments, the freezing point depression of the PVA solutions with various compositions was measured via cooling curve method (see **Figure R2** in **Auxiliary Supplementary Materials 3**). The freezing point depression in PVA solution was expected to be very small due to its high value of the molar mass of PVA chains. And our measurement confirmed this idea in the low concentration range of up to 10 wt.% where a linear relation between freezing point depression and mass fraction of PVA was observed. For higher concentrations exceeding 10 wt.%, non-linear freezing point depression was obtained, which was possibly due to the interaction among PVA chains, swelling and possible gel formation as reported by Kawai [25]. From **Figure R2**, it can be seen that the freezing point depression did not exceed 0.2 K even when the mass percentage reached as high as 10 wt.%, which was consistent with previous reports by Kawai [25] and Blond [26].

### III. Results and discussions

For solutions with low concentrations (0.25 wt.% and 1 wt.%), the S/L interface was always asymmetrically cellular-like with narrow spacing (in **Figure 2** and **Figure 3**). The cellular spacing decreased as the growth velocity increased. The solute accumulation ahead of S/L interface for PVA solution has been reported [43] under lower growth velocity. Then the constitutional undercooling ahead of the S/L interface through accumulation of PVA chains might account for such cellular morphology. The degree of hydrolysis did not make significant differences in morphology evolution of S/L interface for 0.25 wt.% samples as shown in **Figure 2**. But different details in morphology appeared for 1 wt.% samples, where the cellular grooves of PVA203 appeared a little denser than PVA103 as shown in **Figure 3**.

In contrast to the results of low concentration samples, solutions with higher concentrations (5 wt.% and 10 wt.%) presented obvious morphological transition with varying pulling velocity. At the lowest growth velocity regime (3-5 um/s), misty morphology was observed for PVA103 samples (as shown in **Figure 4(a)** and **Figure 5(a)**), in which a series of squared tips could be observed with higher magnification (the inset in **Figure 4(a)**) and there were dense slashes vertical to the direction of pulling velocity. When pulling velocity increased, V-shaped lamellar morphology (as shown in **Figure 4(b-f)** and **Figure 5(b-f)**) appeared on the basis of preformed misty morphology and varied with increasing growth velocity. It can be seen that PVA103 and PVA203 exhibited almost the same tendency of morphological evolution, but PVA203 solutions had not developed misty morphology (see **Figure 6(a)** and **Figure 7(a)**) at the lowest growth velocity in our experiments, which indicated that PVA203 solutions may have a lower transition velocity from misty to V-shaped morphology.

The formation of misty morphology might be related to a mechanical hindrance effect [26, 44] and/or interface attachment processes [21-24] of the accumulated PVA chains. It is reported that [45] the ice nucleation efficiency of PVA is strongly influenced by degree of hydrolysis, which provided useful information of the effect of degree of hydrolysis on the resulting ice morphology. In our study, more freely mobile water molecules is suggested to be present near the S/L interface during directional freezing due to the diminishing constraint by the hydroxyl groups on the PVA chains in solutions [45], which may also qualitatively explain why misty morphology was observed in PVA103 solutions instead of PVA203 under approximately the same low growth velocities. It should be noted here that, few reports can be found on such misty morphology in ice templating.

As for the V-shaped morphology with tilted ice dendrites, a qualitative explanation is given as follows. PVA is widely reported to have antifreeze properties [45-48], which may lead to perturbations in the ice-water interface, allowing ice dendrites to form, possibly related to increased supercooling from rejection of the PVA and interface attachment process of PVA chains on edge plane of ice [47-49]. Because the growth kinetics of edge and basal plane are different due to the different nature of their interfaces, different bath undercoolings will yield different growth rates [50] for edge and basal planes, and the actual growth direction of ice dendritic tip will deviate from basal plane to some extent, depending on the undercooling and solute additives. Therefore, in our study, it is reasonable to speculate that the change of ice dendritic tip growth direction can be qualitatively addressed by the "step growth mechanism". The PVA impurity was suggested to enhance the tilting of cellular patterns, possibly by altering the growth kinetics of basal plane via introducing crystal imperfection as reported by Michaels et al [51], which was also qualitatively consistent with the results of Pruppacher [52] and Macklin & Ryan [50] under free growth conditions.

Studies of microstructure manipulation in ice-templating have been extensively focused because they are directly linked with the mechanical properties of the ice templating samples [53]. In ice templating, the empirical dependence of $\lambda_1$ on $V_{pulling}$ are usually described by a simple power law ($\lambda_1 \propto (\frac{1}{V_{pulling}})^b$) [54-55], which supports the general and intuitive trend that finer lamellar spacing can result from larger pulling velocities. Here, the lamellar spacing $\lambda_1$ was measured optically parallel to cellular grooves and/or lamellar platelets from all PVA samples with obvious cellular and/or lamellar morphology. It should be noted that in this paper, the lamellar platelets were well-defined in terms of the ice crystal orientation and thus provided lamellar spacing which was more well-defined. $\lambda_1$ was plotted against pulling velocity $V_{pulling}$ as shown in **Figure 8**. The plots in **Figure 8** were fitted with power laws with exponent $b$. The data of 1 wt.% PVA103 was absent because of unreliable measurement on its narrow cellular spacing. Different values of exponent $b$ were summarized in **Table 1**. Previous report in PVA/water system [33] yielded similar results with an exponent value of 2/3, which was within the range of this work. It is interesting that $b$ depends on the solute concentration. For all PVA solutions, $b$ increased with PVA concentration, but $b$ seemed to increase faster for PVA203 samples as shown in **Table 1**. This indicated that, in addition to solute concentration, lamellar spacing was also sensitive to the degree of hydrolysis of PVA additives.

Apart from the factors affecting lamellar spacing, the tilted growth of lamellar microstructures has been frequently observed, which reported that lamellar platelets usually formed tilted and crossed domains, and the widely accepted explanation is the combined influence of non-parallel thermal gradient and preferred orientation on actual growth direction of ice dendrites [56-57]. In this paper, however, there was an obvious shift of actual growth direction of lamellar ice tips in the case where the preferred crystal orientation paralleled to the thermal gradient as described in **Figure 1(g-h)**. This is an important challenge to the popular explanation for the formation of tilted lamellar morphology.

This fact could be better elaborated by defining a characteristic tilt angle "$2\theta$"

with respect to the direction of pulling velocity $V_{pulling}$ (as shown in the inset in **Figure 9 (a)**). The variation of the characteristic tilt angle $2\theta$ was measured against increasing pulling velocities $V_{pulling}$ (see **Figure 9 (a)**). It was noticed from **Figure 9 (a)** that when the concentration was lowered from 10 wt.% to 5 wt.%, the average value of $2\theta$ went down to about a third for both PVA103 and PVA203. And it was shown that the degree of hydrolysis of PVA in this work was also a key factor in tuning the characteristic tilt angle $2\theta$. It should be noted that, Pruppacher [52] had summarized many previous results on ice growth habit and obtained two different maximum tilt angles in pure water ($2\theta_{max} = 45°$) and various ionic solutions ($2\theta_{max} = 60°$) with different bath undercoolings of up to 15 K under free growth condition. After reaching a maximum tilt angle, further increment of bath undercooling no longer increased the tilt angle. Although the results in **Figure 9 (a)** did not show an obvious maximum value towards increased growth velocity, the largest tilt angle could reach as large as 90°, which was much larger than previous reports in pure water and ionic solutions. This indicates that PVA may have a great impact on V-shaped lamellar morphology in terms of $2\theta$. Inspired by previous studies on ice growth morphology [50, 52], a schematic description of the mechanism of the tilting was given in **Figure 9 (b)**. In **Figure 9 (b)**, the relation $\vec{V}_{pulling} \parallel \vec{G} \parallel \{0001\}$ is satisfied. Micro-steps may be formed around the ice tip. When $V_{pulling}$ was not large, basal plane grew much slower than edge plane, and no tilting of ice dendrites was present. On the contrary, when $V_{pulling}$ increased to some extent, basal plane may become imperfect due to the presence of PVA chains and increased driving force to freeze and grew at a comparable velocity to that of edge plane, which mae the ice tip distinctly deviate from the direction of thermal gradient.

Tip undercooling (interface undercooling) is usually an important factor in directional solidification, especially for the constitutional undercooling. Accordingly, in ice templating, tip undercooling is of great significance due to its direct link with microscopic pattern formation [14, 58]. In this study, the tip undercoolings $\Delta T_{tip}$ against increasing pulling velocity $V_{pulling}$ under steady state were simultaneously measured and plotted for all samples in **Figure 10**. For dilute solutions (0.25 wt.%, 1 wt.% and 5 wt.%), the results were similar since they yielded an attenuation trend toward a lower limit close to their static interface undercooling. Hence, it is speculated that the tip undercooling is mainly controlled by solute diffusion, which is similar to the solute buildup ahead of the S/L interface at much lower PVA concentration (1 wt.%) as reported by Butler [43]. For the most concentrated samples (10 wt.%), variations of the tip undercooling no longer exhibited attenuation but experienced intermittent elevations which made the tip undercoolings always much higher than those in static case. Such abnormal results in contrast to dilute solutions can be accounted for by concentration-dependent diffusive property of PVA. During directional freezing of concentrated PVA solutions, the possible engulfment and rejection interaction between PVA chains and S/L interface in our study may greatly influence segregation of PVA chains ahead of S/L interface since the diffusion of concentrated PVA chains were severely hindered. Besides, gel formation and chain entanglement of concentrated PVA chains may further lower the diffusivity of PVA chains, affecting the segregation of PVA ahead of S/L interface. There may be an obvious solute buildup layer at increasing pulling velocity for 10 wt.% samples since the tip undercooling is always much higher than static case for all pulling velocities in this study. The complex V-shaped lamellar morphology occurred in such small undercooling range (less than 1.2 K in our study as shown in **Figure 10**) in the presence of concentrated PVA solutions. Owing to this fact, the complex interface morphology evolution may have an inherently different physical origin other than constitutional undercooling.

In addition, it should be noted here that the lamellar microstructure of ice during directional solidification changed with pulling velocity and solute concentration even the preferred crystallographic orientation of ice was fixed parallel to the thermal direction. This is probably one of the key origins of diverse microstructure features during directional ice templating, like ice bridges and randomly orientated lamellar colonies, which has been proved to influence the related properties of porous microstructure significantly. The in-situ quantitative study of S/L interface morphology evolution of ice during directional solidification is helpful to understand and control the sample microstructure fabricated by ice-templating.

## IV. Conclusion

In summary, the S/L interface morphological evolution of a single crystal ice with manipulated orientation was in-situ investigated during directional freezing of PVA solutions with different concentrations and degree of hydrolysis for the first time. Samples with low concentrations (0.25 wt.% and 1 wt.%) could not develop obvious lamellar morphology in this study. In contrast, samples with high concentrations (5 wt.% and 10 wt.%) experienced a morphological evolution from adsorption-controlled misty morphology to kinetically-controlled V-shaped lamellar morphology. Based on the measured lamellar spacing and tilt angle of ice platelets of V-shaped lamellar morphology, it was found that PVA103 was more effective in increasing the characteristic tilt angle of V-shaped lamellar morphology than PVA203 as the growth velocity increased. The variation of the measured tip undercooling against growth velocity of concentrated PVA samples showed unusual behavior, which indicated a complex interaction between the concentrated polymer chains and the S/L interface. Based on the results, it can be concluded that apart from the mass fraction and size effect of solute additives, the growth morphology of ice in ice templating can also be significantly tuned by the degree of hydrolysis of concentrated polymer additives. More delicate and quantitative studies are needed to explore the nature of directional solidification of polymer solutions, which is believed to influence future design and optimization in ice-templating.

## Acknowledgements

This work was supported by National Natural Science Foundation of China (Grant No. 51701155), the National Key R&D Program of China (Grant No.2018YFB1106003) and the Fundamental Research Funds for the Central Universities (3102019ZD0402).

**Reference:**

1. Shao, G.; Hanaor, D. A. H.; Shen, X.; Gurlo, A., Freeze Casting: From Low-Dimensional Building Blocks to Aligned Porous Structures—A Review of Novel Materials, Methods, and Applications. *Advanced Materials* **2020,** *32* (17), 1907176.

2. Shaga, A.; Shen, P.; Guo, R.-F.; Jiang, Q.-C., Effects of oxide addition on the microstructure and mechanical properties of lamellar SiC scaffolds and Al–Si–Mg/SiC composites prepared by freeze casting and pressureless infiltration. *Ceramics International* **2016,** *42* (8), 9653-9659.

3. Huang, J.; Xu, Z.; Moreno, S.; Morsali, S.; Zhou, Z.; Daryadel, S.; Baniasadi, M.; Qian, D.; Minary‐Jolandan, M., Lamellar Ceramic Semicrystalline‐Polymer Composite Fabricated by Freeze Casting. *Advanced Engineering Materials* **2017,** *19* (8), 1700214.

4. Wu, T.; Zhang, W.; Yu, B.; Chen, J., A novel electrolyte-electrode interface structure with directional micro-channel fabricated by freeze casting: A minireview. *International Journal of Hydrogen Energy* **2017,** *42* (50), 29900-29910.

5. Yu, Z.-L.; Yang, N.; Zhou, L.-C.; Ma, Z.-Y.; Zhu, Y.-B.; Lu, Y.-Y.; Qin, B.; Xing, W.-Y.; Ma, T.; Li, S.-C., Bioinspired polymeric woods. *Science advances* **2018,** *4* (8), eaat7223.

6.  Xu, W.; Xing, Y.; Liu, J.; Wu, H.; Cui, Y.; Li, D.; Guo, D.; Li, C.; Liu, A.; Bai, H., Efficient water transport and solar steam generation via radially, hierarchically structured aerogels. *ACS nano* **2019,** *13* (7), 7930-7938.

7.  Bai, H.; Walsh, F.; Gludovatz, B.; Delattre, B.; Huang, C.; Chen, Y.; Tomsia, A.; Ritchie, R., Bioinspired Hydroxyapatite/Poly(Methyl Methacrylate) Composite with Nacre-Mimetic Architecture by a Bidirectional Freezing Method. *Advanced Materials* **2015**.

8.  Bai, H.; Chen, Y.; Delattre, B.; Tomsia, A.; Ritchie, R., Bioinspired Large-Scale Aligned Porous Materials Assembled with Dual Temperature Gradients. *Science Advances* **2015,** *1*, e1500849.

9.  Zhang, H.; Hussain, I.; Brust, M.; Butler, M.; Rannard, S.; Cooper, A., Aligned Two- and Three-Dimensional Freezing of Polymers and Nanoparticles. *Nature materials* **2005,** *4*, 787-93.

10. Donius, A. E.; Obbard, R. W.; Burger, J. N.; Hunger, P. M.; Baker, I.; Doherty, R. D.; Wegst, U. G. K., Cryogenic EBSD reveals structure of directionally solidified ice–polymer composite. *Materials Characterization* **2014,** *93*, 184-190.

11. Kurz, W.; Fisher, D. J.; Trivedi, R., Progress in modelling solidification microstructures in metals and alloys: dendrites and cells from 1700 to 2000. *International Materials Reviews* **2019,** *64* (6), 311-354.

12. Kurz, W.; Rappaz, M.; Trivedi, R., Progress in modelling solidification microstructures in metals and alloys. Part II: dendrites from 2001 to 2018. *International Materials Reviews* **2020**, 1-47.

13. You, J.; Wang, L.; Wang, Z.; Li, J.; Wang, J.; Lin, X.; Huang, W., In situ observation the interface undercooling of freezing colloidal suspensions with differential visualization method. *Review of Scientific Instruments* **2015,** *86* (8), 084901.

14. You, J.; Wang, L.; Zhijun, W.; Li, J.; Wang, J.; Lin, X.; Huang, W., Interfacial undercooling in solidification of colloidal suspensions: Analyses with quantitative measurements. *Scientific Reports* **2016,** *6*, 28434.

15. Wang, L.; You, J.; Zhijun, W.; Wang, J.; Lin, X., Interface instability modes in freezing colloidal suspensions: Revealed from onset of planar instability. *Scientific reports* **2015,** *6*.

16. Doi, M., *Soft Matter Physics*. Oxford University Press: Oxford, United Kingdom, 2013.

17. Petit, J.-M.; Roux, B.; Zhu, X.; Macdonald, P., A new physical model for the diffusion of solvents and solute probes in polymer solutions. *Macromolecules* **1996,** *29* (18), 6031-6036.

18. Vrentas, J.; Duda, J.; Ni, L., Concentration dependence of polymer self-diffusion coefficients. *Macromolecules* **1983,** *16* (2), 261-266.

19. Teraoka, I., *Polymer solutions: an introduction to physical properties*.

John Wiley & Sons: 2002.

20. Masaro, L.; Zhu, X., Physical models of diffusion for polymer solutions, gels and solids. *Progress in polymer science* **1999,** *24* (5), 731-775.

21. Wang, H.-Y.; Inada, T.; Funakoshi, K.; Lu, S.-S., Inhibition of nucleation and growth of ice by poly (vinyl alcohol) in vitrification solution. *Cryobiology* **2009,** *59* (1), 83-89.

22. Chasnitsky, M.; Braslavsky, I., Ice-binding proteins and the applicability and limitations of the kinetic pinning model. *Philosophical Transactions of the Royal Society A* **2019,** *377* (2146), 20180391.

23. Liu, K.; Wang, C.; Ma, J.; Shi, G.; Yao, X.; Fang, H.; Song, Y.; Wang, J., Janus effect of antifreeze proteins on ice nucleation. *Proceedings of the National Academy of Sciences* **2016,** *113* (51), 14739-14744.

24. Yeh, Y.; Feeney, R. E., Antifreeze proteins: structures and mechanisms of function. *Chemical reviews* **1996,** *96* (2), 601-618.

25. Kawai, T., Freezing Point Depression of Polymer Solutions and Gels. *Journal of Polymer Science* **1958,** *32* (125), 425-444.

26. Blond, G., Freezing in Polymer — Water Systems and Properties of Water. In *Properties of Water in Foods*, Springer, Dordrecht: 1985; Vol. 90, pp 531-542.

27. Brewster; D., On the Affections of Light Transmitted through Crystallized Bodies. *Philosophical Transactions of the Royal Society of*

*London* **1814,** *104*, 187-218.

28. Brewster; D., On the Laws of Polarisation and Double Refraction in Regularly Crystallized Bodies. *Philosophical Transactions of the Royal Society of London* **1818**.

29. Petrenko, V.; Whitworth, R., *Physics of ice*. Oxford University Press: 2002.

30. Zhang, T. X.; Wang, Z. J.; Wang, L. L.; Li, J. J.; Lin, X.; Wang, J. C., Orientation determination and manipulation of single ice crystal via unidirectional solidification. *Acta Physica Sinica* **2018,** *67* (19).

31. Peko, C.; Groth, B.; Nettleship, I., The effect of polyvinyl alcohol on the microstructure and permeability of freeze‐cast alumina. *Journal of the American Ceramic Society* **2010,** *93* (1), 115-120.

32. Zuo, K. H.; Zeng, Y.-P.; Jiang, D., Effect of polyvinyl alcohol additive on the pore structure and morphology of the freeze-cast hydroxyapatite ceramics. *Materials Science and Engineering: C* **2010,** *30* (2), 283-287.

33. Zhang, H.; Hussain, I.; Brust, M.; Butler, M. F.; Rannard, S. P.; Cooper, A. I., Aligned two-and three-dimensional structures by directional freezing of polymers and nanoparticles. *Nature materials* **2005,** *4* (10), 787-793.

34. Gutiérrez, M.; Garcia, Z.; Jobbagy, M.; Rubio, F.; Yuste, L.; Rojo, F.; Ferrer, M. L.; Monte, F., Poly(vinyl alcohol) Scaffolds with Tailored Morphologies for Drug Delivery and Controlled Release. *Advanced*

*Functional Materials* **2007,** *17*, 3505-3513.

35. Schoof, H.; Bruns, L.; Fischer, A.; Heschel, I.; Rau, G., Dendritic ice morphology in unidirectionally solidified collagen suspensions. *Journal of crystal growth* **2000,** *209* (1), 122-129.

36. Naullage, P. M.; Lupi, L.; Molinero, V., Molecular recognition of ice by fully flexible molecules. *The Journal of Physical Chemistry C* **2017,** *121* (48), 26949-26957.

37. Holloway, J. L.; Lowman, A. M.; Palmese, G. R., The role of crystallization and phase separation in the formation of physically cross-linked PVA hydrogels. *Soft Matter* **2013,** *9* (3), 826-833.

38. Komatsu, M.; Inoue, T.; Miyasaka, K., Light‑scattering studies on the sol–gel transition in aqueous solutions of poly (vinyl alcohol). *Journal of Polymer Science Part B: Polymer Physics* **1986,** *24* (2), 303-311.

39. Pekor, C.; Nettleship, I., The effect of the molecular weight of polyethylene glycol on the microstructure of freeze-cast alumina. *Ceramics International* **2014,** *40*, 9171-9177.

40. Peko, C.; Groth, B.; Nettleship, I., The Effect of Polyvinyl Alcohol on the Microstructure and Permeability of Freeze‑Cast Alumina. *Journal of the American Ceramic Society* **2010,** *93*, 115-120.

41. Porter, M.; Imperio, R.; Wen, M.; McKittrick, J., Bioinspired Scaffolds with Varying Pore Architectures and Mechanical Properties. *Advanced Functional Materials* **2014,** *24*.

42. Zuo, K.; Zeng, Y.-P.; Jiang, D., Effect of polyvinyl alcohol additive on the pore structure and morphology of the freeze-cast hydroxyapatite ceramics. *Materials Science and Engineering: C* **2010,** *30*, 283-287.

43. Butler, M. F., Freeze concentration of solutes at the ice/solution interface studied by optical interferometry. *Crystal growth & design* **2002,** *2* (6), 541-548.

44. Blond, G., Velocity of linear crystallization of ice in macromolecular systems. *Cryobiology* **1988,** *25*, 61-6.

45.Wu, S.; He, Z.; Zang, J.; Jin, S.; Wang, Z.; Wang, J.; Yao, Y.; Wang, J., Heterogeneous ice nucleation correlates with bulk-like interfacial water. *Science Advances* **2019,** *5*, eaat9825.

46. Wang, H.-Y.; Inada, T.; Funakoshi, K.; Lu, S.-S., Inhibition of nucleation and growth of ice by poly(vinyl alcohol) in vitrification solution. *Cryobiology* **2009,** *59*, 83-89.

47. Lupi, L.; Molinero, V., Molecular Recognition of Ice by Fully Flexible Molecules. *The Journal of Physical Chemistry C* **2017,** *121*.

48. Budke, C.; Koop, T., Ice Recrystallization Inhibition and Molecular Recognition of Ice Faces by Poly(vinyl alcohol). *ChemPhysChem* **2006,** *7* (12), 2601-2606.

49. Qiu, Y.; Molinero, V., What Controls the Limit of Supercooling and Superheating of Pinned Ice Surfaces? *The Journal of Physical Chemistry Letters* **2018,** *9*.

50. Macklin, W.; Ryan, B., Growth velocities of ice in supercooled water and aqueous sucrose solutions. *The Philosophical Magazine: A Journal of Theoretical Experimental and Applied Physics* **1968,** *17* (145), 83-87.

51. Michaels, A.; Brian, P.; Sperry, P., Impurity Effects on the Basal Plane Solidification Kinetics of Supercooled Water. *Journal of Applied Physics* **1967,** *37*, 4649-4661.

52. Pruppacher, H., Growth Modes of Ice Crystals in Supercooled Water and Aqueous Solutions. *Journal of Glaciology* **1967,** *6*, 651-662.

53. Deville, S.; Saiz, E.; Nalla, R. K.; Tomsia, A. P., Freezing as a path to build complex composites. *Science* **2006,** *311* (5760), 515-518.

54. Waschkies, T.; Oberacker, R.; Hoffmann, M., Investigation of structure formation during freeze-casting from very slow to very fast solidification velocities. *Acta Materialia* **2011,** *59* (13), 5135-5145.

55. Waschkies, T.; Oberacker, R.; Hoffmann, M. J., Control of lamellae spacing during freeze casting of ceramics using double-side cooling as a novel processing route. *Journal of the American Ceramic Society* **2009,** *92*, S79-S84.

56. Deville, S.; Saiz, E.; Tomsia, A. P., Ice-templated porous alumina structures. *Acta Materialia* **2007,** *55* (6), 1965-1974.

57. Dedovets, D.; Deville, S., Multiphase imaging of freezing particle suspensions by confocal microscopy. *Journal of the European Ceramic Society* **2018,** *38* (7), 2687-2693.

58. Peppin, S. S. L.; Wettlaufer, J. S.; Worster, M. G., Experimental Verification of Morphological Instability in Freezing Aqueous Colloidal Suspensions. *Physical Review Letters* **2008**, *100* (23), 238301.

**Figure captions:**

**Figure 1** The schematic description of orientation manipulation of ice crystal and S/L interface undercooling measurement by a redesigned horizontal Bridgeman apparatus. **Figure** (**a**-**c**) are the physical foundation of the orientation detection based on crystal optics. **Figure** (**d**-**g**) are the specific operations for manipulating a single ice crystal with designed orientation in a series of glass specimen boxes. **Figure** (**h**) is the principle for tip undercooling measurement. (**a**) The orientation relation between the single crystal ice in a glass specimen box tied to the frame ***X-Y-Z*** and the laboratory frame ***A-P-L*** with two parameters α and β; (**b**) $\Delta N_\alpha$-α curve, the position of the maximum $\Delta N_\alpha$ satisfies α = 90°, labeled as "*M*"; (**c**) The dimensionless intensity $I_\perp/I_0$ which corresponds to the length of line "*OF*" against extinction angle β on polar coordinate system, the complete extinction positions are labeled as "*E1*"(β = 0°) and "*E2*"(β = 90°), respectively; (**d**)The orientation relation of the single ice crystal grown in the first specimen box (the frame ***X$_0$-Y$_0$-Z$_0$***) in the frame ***A-P-L*** is $\alpha_0$, $\beta_0$; (**e**)The orientation relation between the first (on the left) and the second (on the right) specimen boxes, with the projection of optical axis of ice parallel to ***P-P*** direction by rotating the first specimen box to position "*E1*" so that the orientation relation of the single ice crystal in the second specimen box (the frame ***X$_1$-Y$_1$-Z$_1$***) in the frame ***A-P-L*** is $\alpha_1 = \alpha_0$, $\beta_1 = 0°$; (**f**)The optical axis of ice is made to lie in the plane ***A-P*** by putting down the third specimen box so that the orientation relation of the single ice crystal in the third specimen box (the frame ***X$_2$-Y$_2$-Z$_2$***) in the frame ***A-P-L*** is $\alpha_2 = 90°$ (position "*M*"), $\beta_1 = 90° - \alpha_0$; (**g**)The orientation of each ice crystal grown in capillary tubes in this paper satisfies the positions "*M*" and "*E1*" as shown in the fourth specimen box; (**h**)Schematic diagram of horizontal directional freezing stage and measurement of tip

undercooling $\Delta T_{tip}$ by differential visualization method. The crystal orientations of ice crystals in two glass tubes for directional growth are the same as represented by the red solid rod lying within the plane *X-Y* and parallel to the *Y* axis in which the laboratory frame *A-P-L* coincides with the frame $X_3$-$Y_3$-$Z_3$. The temperatures of both heating and cooling zones are provided by ethanol thermostat whose temperature can change from −20.0 ∘C to 50.0 ∘C discretionarily. Both heaters and coolers are made of copper blocks (120 × 100 × 10 mm) with ethanol circulation. The temperature of the thermostat was set by a temperature controller. Sample translation across the thermal gradient is provided by a servo motor supplemented with a linear ball-screw drive (labeled as "*M*" in **Figure** (**h**)), allowing the growth rate of ice samples to be changed. Observation of the growth front is achieved through an optical microscope stage with a charge-coupled device (CCD) camera.

**Figure 2** Microstructure evolution of 0.25 wt. % PVA103 (**a-c**) and PVA203 (**d-f**) solutions under increasing growth velocities. The growth velocities for (**a**)-(**f**) were 3.95 um/s, 19.47 um/s, 32.10 um/s, 4.13 um/s, 19.27 um/s and 35.65 um/s, respectively. The temperature gradients for the PVA103 and PVA203 samples were G = 5.8 K/mm and 4.3 K/mm. The pulling velocity was parallel to both thermal gradient and basal plane {0001} of ice.

**Figure 3** Microstructure evolution of 1 wt. % PVA103 (**a-c**) and PVA203 (**d-f**) solutions under increasing growth velocities. The growth velocities for (**a**)-(**f**) were 3.86 um/s, 19.36 um/s, 36.07 um/s, 4.10 um/s, 19.30 um/s and 35.65 um/s, respectively. The temperature gradients for the PVA103 and PVA203 samples were G = 4.6 K/mm and 5.1 K/mm. The microstructures for PVA203 at each growth velocity yielded more cellular grooves than PVA103. The pulling velocity was parallel to both thermal gradient and basal plane {0001} of ice.

**Figure 4** Microstructure evolution of 5 wt. % PVA103 solutions under increasing growth velocities. Two strokes in magenta in (**f**) showed the formation of V-shaped

lamellar platelets. The growth velocities for (**a**)-(**f**) were 4.15 um/s, 8.68 um/s, 13.31 um/s, 25.79 um/s, 36.93 um/s and 56.28 um/s, respectively. The temperature gradient for the sample was G = 4.8 K/mm. The pulling velocity was parallel to both thermal gradient and basal plane {0001} of ice.

**Figure 5** Microstructure evolution of 10 wt. % PVA103 solutions under increasing growth velocities. The growth velocities for (**a**)-(**f**) were 3.98 um/s, 8.30 um/s, 13.31 um/s, 24.84 um/s, 35.91 um/s and 52.86 um/s, respectively. The temperature gradient for the sample was G = 5.5 K/mm. The pulling velocity was parallel to both thermal gradient and basal plane {0001} of ice.

**Figure 6** Microstructure evolution of 5 wt. % PVA203 solutions under increasing growth velocities. The growth velocities for (**a**)-(**f**) were 4.15 um/s, 8.06 um/s, 12.89 um/s, 25.08 um/s, 35.62 um/s and 54.53 um/s, respectively. The temperature gradient for the sample was G = 4.9 K/mm. The pulling velocity was parallel to both thermal gradient and basal plane {0001} of ice.

**Figure 7** Microstructure evolution of 10 wt. % PVA203 solutions under increasing growth velocities. The growth velocities for (**a**)-(**f**) were 4.12 um/s, 8.44 um/s, 13.31 um/s, 25.44 um/s, 35.65 um/s and 53.76 um/s, respectively. The temperature gradient for the sample was G = 5.3 K/mm. The pulling velocity was parallel to both thermal gradient and basal plane {0001} of ice.

**Figure 8** Variation of primary lamellar spacing $\lambda_1$ against pulling velocity $V_{pulling}$ during steady-state freezing for all PVA samples in this study. The data points for each PVA sample were fitted with the power law $\lambda_1 \propto (\dfrac{1}{V_{pulling}})^b$.

**Table 1** Determined power law $\lambda_1 \propto (\dfrac{1}{V_{pulling}})^b$ exponents $b$ for all PVA samples based on the results of **Figure 8**.

**Figure 9** (**a**) Variation of the characteristic tilt angle 2θ against increasing pulling velocities with two concentrations of 5 wt.% and 10 wt.% for PVA103 and PVA203. The inset showed the definition of half of "characteristic tilt angle 2θ" with respect to pulling velocity $\vec{V}_{pulling}$, ice lamellar tip growth velocity $\vec{V}_{tip}$ for the 10 wt.% PVA203 sample at a pulling velocity of 4.12 um/s and a temperature gradient of G = 5.3 K/mm, in which the temperature gradient is parallel to the directions of both pulling velocity and basal plane $\{0001\}$ of ice. (**b**) Schematic diagram of step growth mechanism of directionally solidified ice platelet ($\vec{G}$ is the thermal gradient that parallels to basal plane $\{0001\}$ and pulling velocity $\vec{V}_{pulling}$) under increased pulling velocities ($V_1 < V_2 < V_3$). The dash line represents the ice tip morphology as a result of directional growth in two perpendicular directions. The solid line represents micro-steps along the ice tip due to ice growth in basal and edge planes. The inset consisting of three arrows with a half of characteristic tilt angle θ shows that the ice lamellar tip growth velocity $\vec{V}_{tip}$ is the combined result of growth velocities in both basal and edge plane directions. PVA impurity is speculated to affect basal plane growth kinetics. Tilting of ice lamellar tip is influenced by both PVA impurity and $\vec{V}_{pulling}$.

**Figure 10** Tip undercooling of all PVA samples against increasing pulling velocities under steady state.

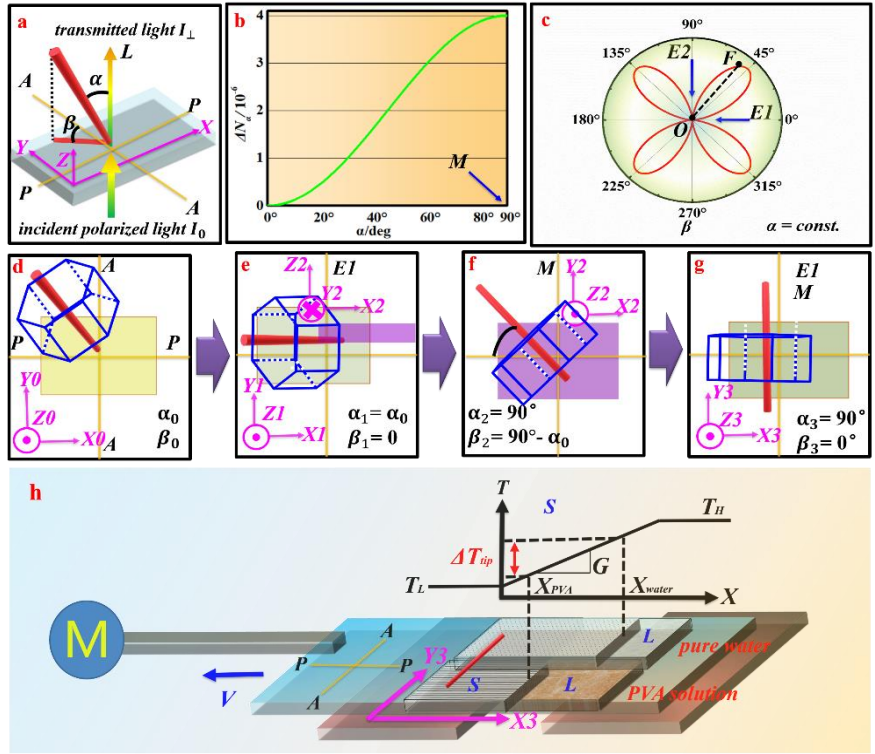

**Figure 1** The schematic description of orientation manipulation of ice crystal and S/L interface undercooling measurement by a redesigned horizontal Bridgeman apparatus. **Figure (a-c)** are the physical foundation of the orientation detection based on crystal optics. **Figure (d-g)** are the specific operations for manipulating a single ice crystal with designed orientation in a series of glass specimen boxes. **Figure (h)** is the principle for tip undercooling measurement. (**a**) The orientation relation between the single crystal ice in a glass specimen box tied to the frame **X-Y-Z** and the laboratory frame **A-P-L** with two parameters α and β; (**b**) $\Delta N_{\alpha}$-α curve, the position of the maximum $\Delta N_{\alpha}$ satisfies α = 90°, labeled as "*M*"; (**c**) The dimensionless intensity $I_{\perp}/I_0$ which corresponds to the length of line "*OF*" against extinction angle β on polar coordinate system, the complete extinction positions are labeled as "*E1*"(β = 0°) and "*E2*"(β = 90°), respectively; (**d**)The orientation relation of the single ice crystal grown in the first specimen box (the frame $X_0$-$Y_0$-$Z_0$) in the frame **A-P-L** is $\alpha_0$, $\beta_0$; (**e**)The orientation relation between the first (on the left) and the second (on the right) specimen boxes, with the projection of optical axis of ice parallel to **P-P** direction by rotating the first specimen box to position "*E1*" so that the orientation relation of the single ice

crystal in the second specimen box (the frame $X_1$-$Y_1$-$Z_1$) in the frame $A$-$P$-$L$ is $\alpha_1 = \alpha_0$, $\beta_1 = 0°$; (**f**)The optical axis of ice is made to lie in the plane $A$-$P$ by putting down the third specimen box so that the orientation relation of the single ice crystal in the third specimen box (the frame $X_2$-$Y_2$-$Z_2$) in the frame $A$-$P$-$L$ is $\alpha_2 = 90°$ (position "$M$"), $\beta_1 = 90°$- $\alpha_0$; (**g**)The orientation of each ice crystal grown in capillary tubes in this paper satisfies the positions "$M$" and "$E1$" as shown in the fourth specimen box; (**h**)Schematic diagram of horizontal directional freezing stage and measurement of tip undercooling $\Delta T_{tip}$ by differential visualization method. The crystal orientations of ice crystals in two glass tubes for directional growth are the same as represented by the red solid rod lying within the plane $X$-$Y$ and parallel to the $Y$ axis in which the laboratory frame $A$-$P$-$L$ coincides with the frame $X_3$-$Y_3$-$Z_3$. The temperatures of both heating and cooling zones are provided by ethanol thermostat whose temperature can change from $-20.0$ ℃ to $50.0$ ℃ discretionarily. Both heaters and coolers are made of copper blocks ($120 \times 100 \times 10$ mm) with ethanol circulation. The temperature of the thermostat was set by a temperature controller. Sample translation across the thermal gradient is provided by a servo motor supplemented with a linear ball-screw drive (labeled as "$M$" in **Figure** (**h**)), allowing the growth rate of ice samples to be changed. Observation of the growth front is achieved through an optical microscope stage with a charge-coupled device (CCD) camera.

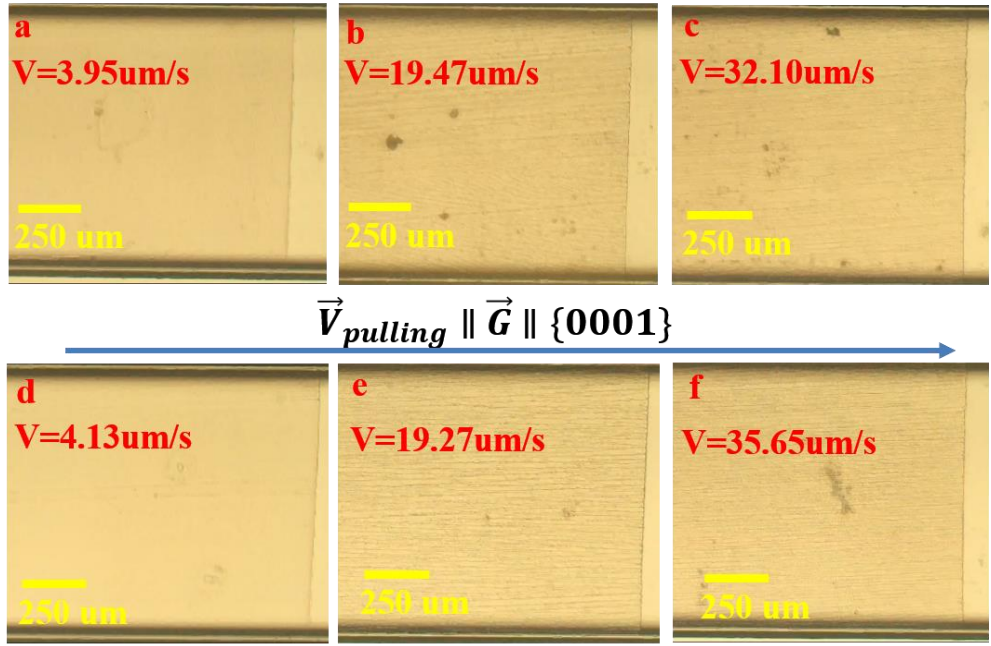

**Figure 2** Microstructure evolution of 0.25 wt. % PVA103 (**a-c**) and PVA203 (**d-f**) solutions under increasing growth velocities. The growth velocities for (**a**)-(**f**) were 3.95 um/s, 19.47 um/s, 32.10 um/s, 4.13 um/s, 19.27 um/s and 35.65 um/s, respectively. The temperature gradients for the PVA103 and PVA203 samples were G = 5.8 K/mm and 4.3 K/mm. The pulling velocity was parallel to both thermal gradient and basal plane {0001} of ice.

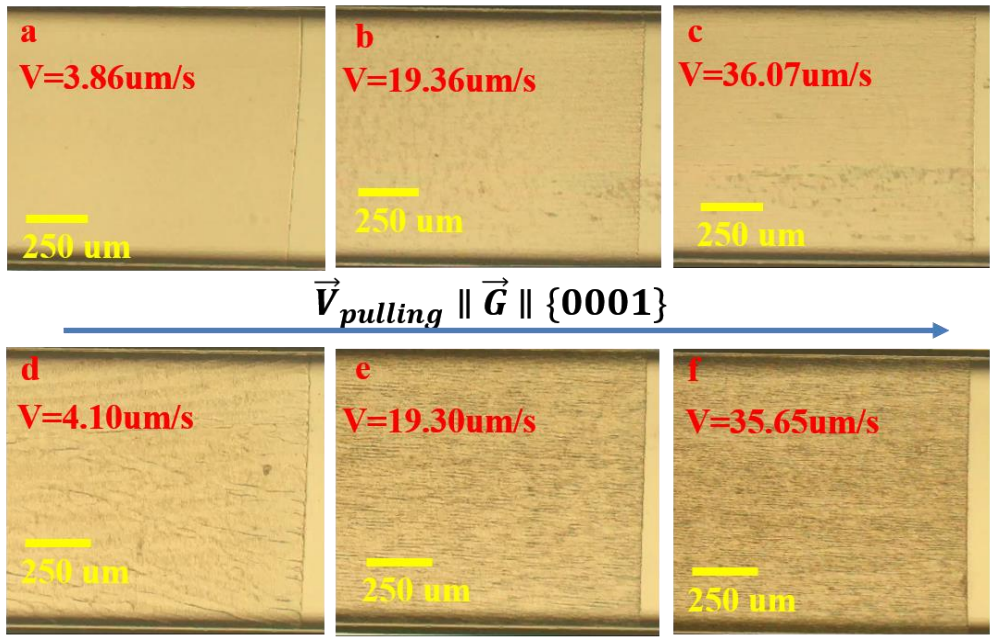

**Figure 3** Microstructure evolution of 1 wt. % PVA103 (**a-c**) and PVA203 (**d-f**) solutions under increasing growth velocities. The growth velocities for (**a**)-(**f**) were 3.86 um/s, 19.36 um/s, 36.07 um/s, 4.10 um/s, 19.30 um/s and 35.65 um/s, respectively. The temperature gradients for the PVA103 and PVA203 samples were G = 4.6 K/mm and 5.1 K/mm. The microstructures for PVA203 at each growth velocity yielded more cellular grooves than PVA103. The pulling velocity was parallel to both thermal gradient and basal plane {0001} of ice.

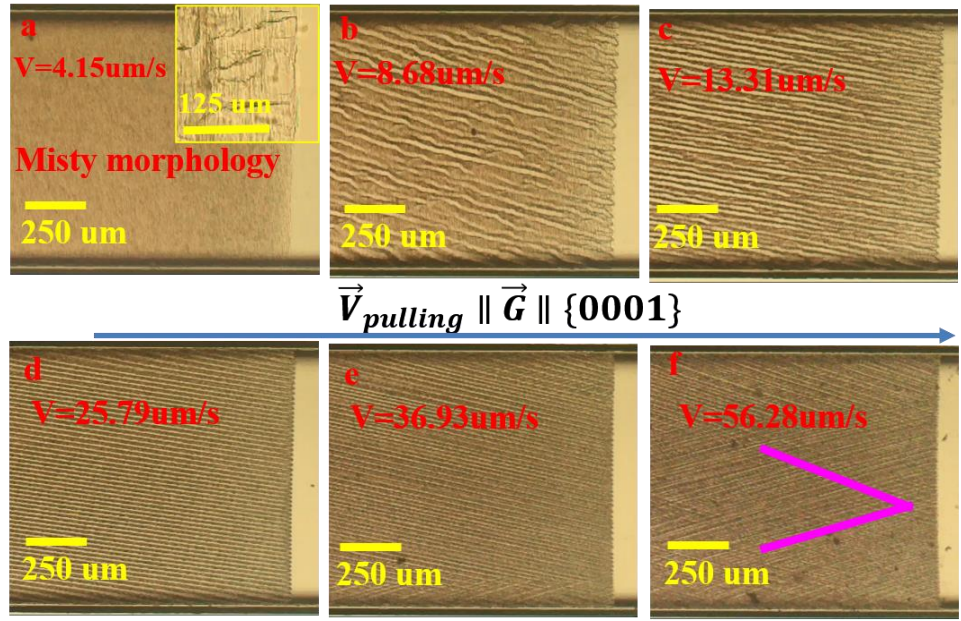

**Figure 4** Microstructure evolution of 5 wt. % PVA103 solutions under increasing growth velocities. Two strokes in magenta in (**f**) showed the formation of V-shaped lamellar platelets. The growth velocities for (**a**)-(**f**) were 4.15 um/s, 8.68 um/s, 13.31 um/s, 25.79 um/s, 36.93 um/s and 56.28 um/s, respectively. The temperature gradient for the sample was G = 4.8 K/mm. The pulling velocity was parallel to both thermal gradient and basal plane {0001} of ice.

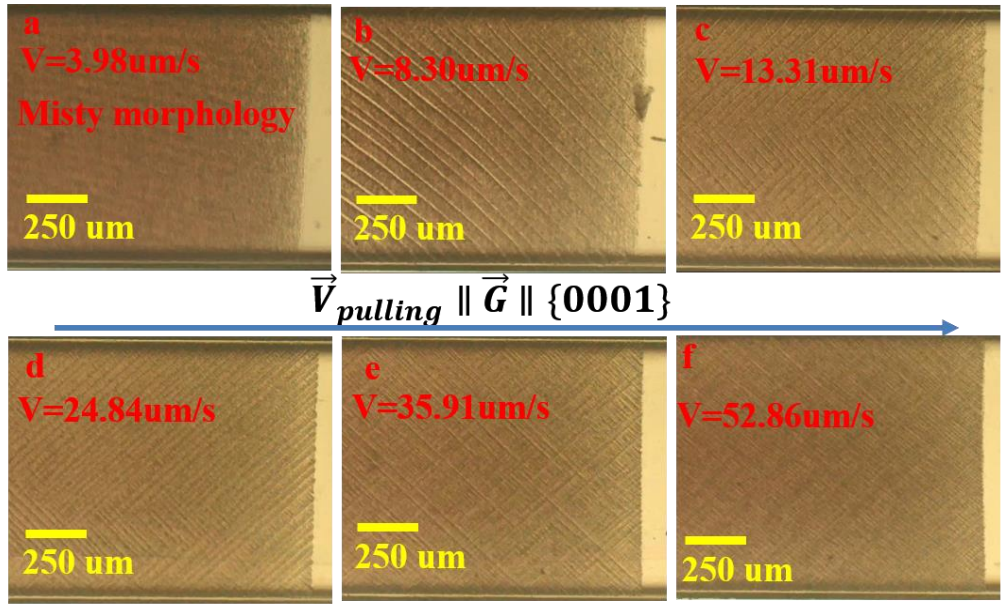

**Figure 5** Microstructure evolution of 10 wt. % PVA103 solutions under increasing growth velocities. The growth velocities for (**a**)-(**f**) were 3.98 um/s, 8.30 um/s, 13.31 um/s, 24.84 um/s, 35.91 um/s and 52.86 um/s, respectively. The temperature gradient for the sample was G = 5.5 K/mm. The pulling velocity was parallel to both thermal gradient and basal plane {0001} of ice.

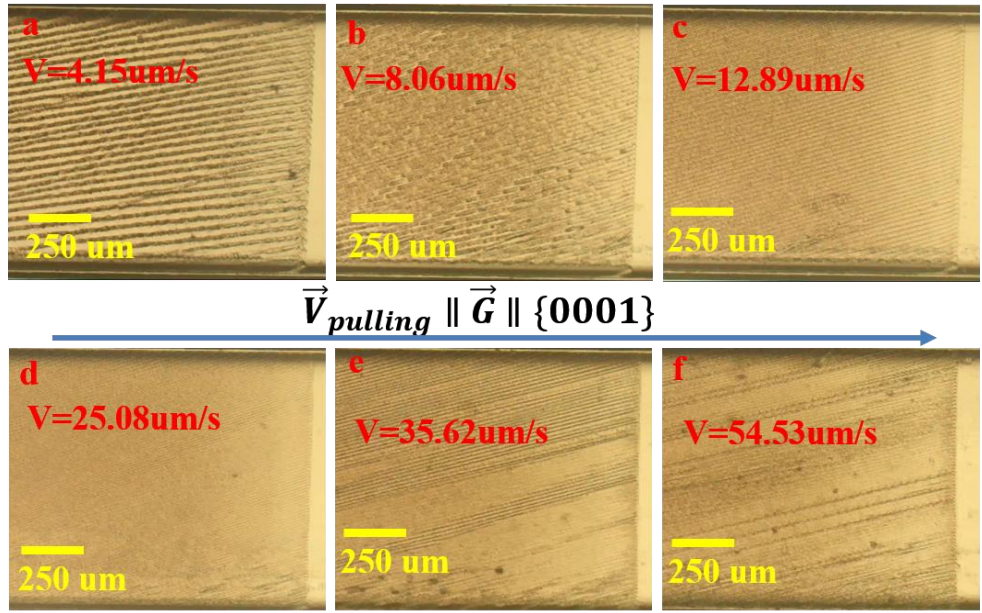

**Figure 6** Microstructure evolution of 5 wt. % PVA203 solutions under increasing growth velocities. The growth velocities for (**a**)-(**f**) were 4.15 um/s, 8.06 um/s, 12.89 um/s, 25.08 um/s, 35.62 um/s and 54.53 um/s, respectively. The temperature gradient for the sample was G = 4.9 K/mm. The pulling velocity was parallel to both thermal gradient and basal plane {0001} of ice.

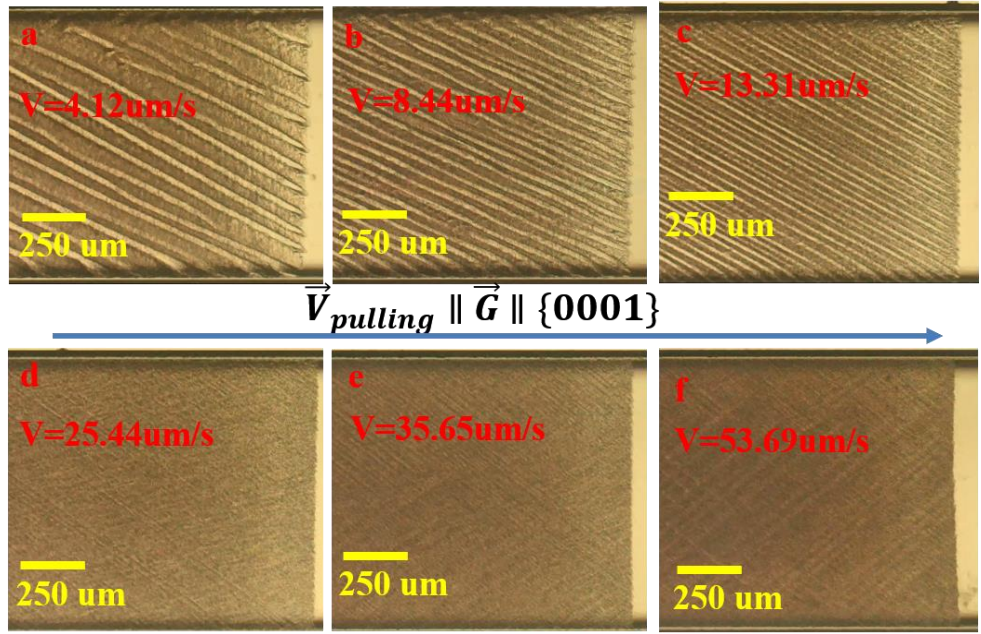

$$\vec{V}_{pulling} \parallel \vec{G} \parallel \{0001\}$$

**Figure 7** Microstructure evolution of 10 wt. % PVA203 solutions under increasing growth velocities. The growth velocities for (**a**)-(**f**) were 4.12 um/s, 8.44 um/s, 13.31 um/s, 25.44 um/s, 35.65 um/s and 53.76 um/s, respectively. The temperature gradient for the sample was G = 5.3 K/mm. The pulling velocity was parallel to both thermal gradient and basal plane {0001} of ice.

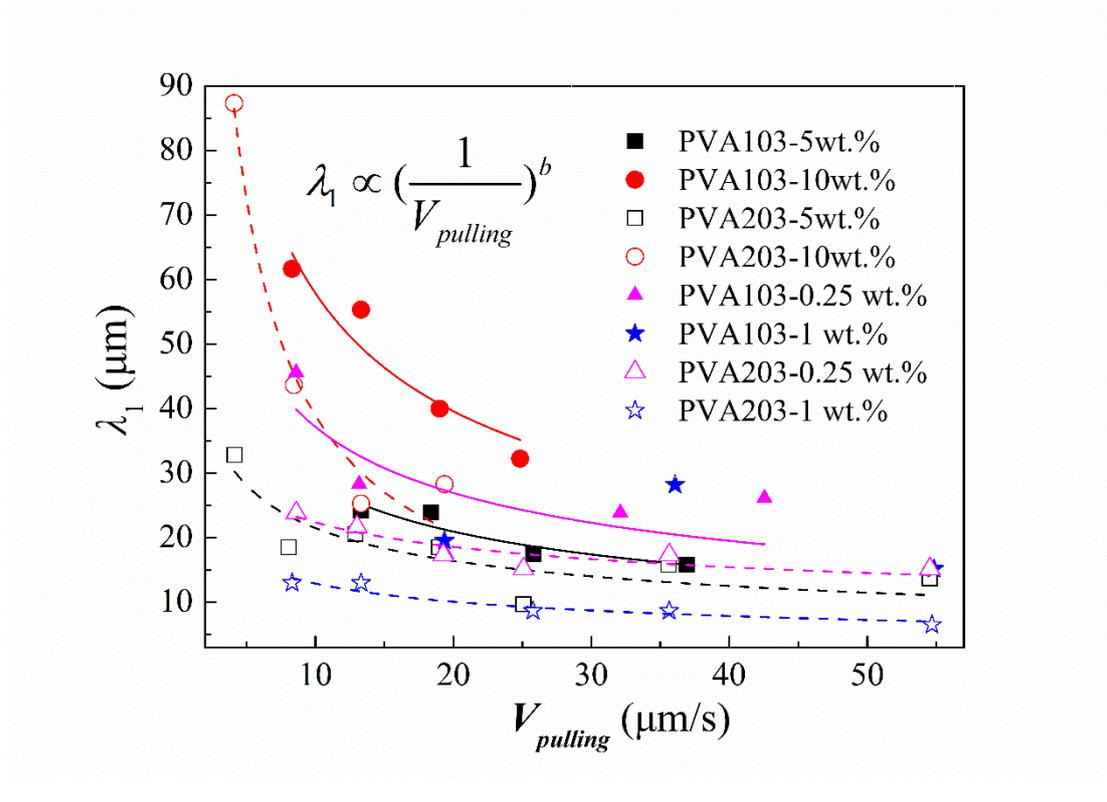

**Figure 8** Variation of primary lamellar spacing $\lambda_1$ against pulling velocity $V_{pulling}$ during steady-state freezing for all PVA samples in this study. The data points for each PVA sample were fitted with the power law $\lambda_1 \propto (\frac{1}{V_{pulling}})^b$.

**Table 1** Determined power law $\lambda_1 \propto (\frac{1}{V_{pulling}})^b$ exponents $b$ for all PVA samples based on the results of **Figure 8**.

| systems | PVA concentration (wt. %) | $b$ |
|---|---|---|
| PVA103/water | 0.25 | 0.464 |
| | 5 | 0.456 |
| | 10 | 0.548 |
| PVA203/water | 0.25 | 0.267 |
| | 1 | 0.356 |
| | 5 | 0.388 |
| | 10 | 0.899 |
| PVA/water [33] | 5 | 0.667 |

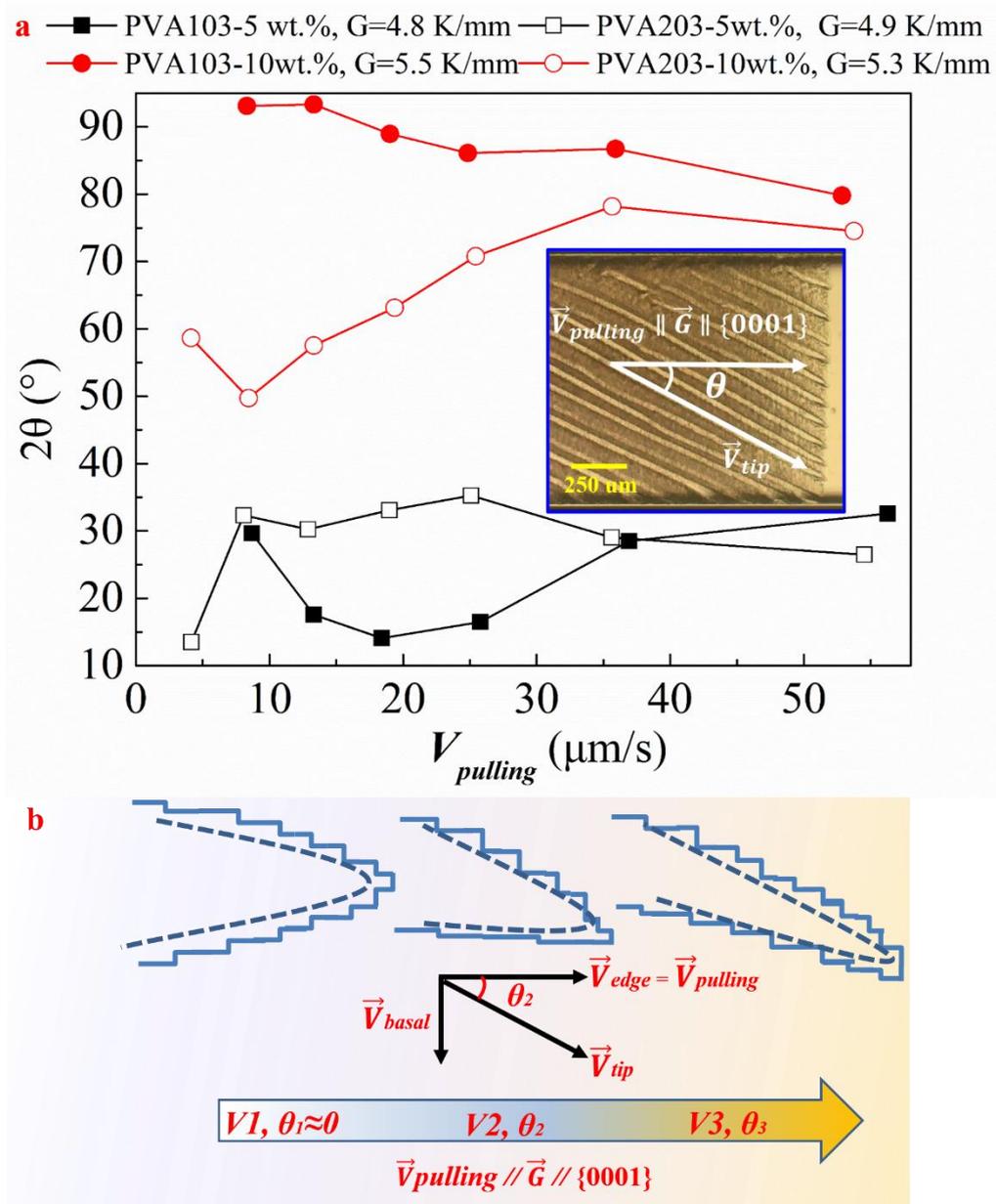

**Figure 9** (**a**) Variation of the characteristic tilt angle 2θ against increasing pulling velocities with two concentrations of 5 wt.% and 10 wt.% for PVA103 and PVA203. The inset showed the definition of half of "characteristic tilt angle 2θ" with respect to pulling velocity $\vec{V}_{pulling}$, ice lamellar tip growth velocity $\vec{V}_{tip}$ for the 10 wt.% PVA203 sample at a pulling velocity of 4.12 um/s and a temperature gradient of G = 5.3 K/mm, in which the temperature gradient is parallel to the directions of both pulling velocity and basal plane {0001} of ice. (**b**) Schematic diagram of step

growth mechanism of directionally solidified ice platelet ($\vec{G}$ is the thermal gradient that parallels to basal plane {0001} and pulling velocity $\vec{V}_{pulling}$) under increased pulling velocities ($V_1 < V_2 < V_3$). The dash line represents the ice tip morphology as a result of directional growth in two perpendicular directions. The solid line represents micro-steps along the ice tip due to ice growth in basal and edge planes. The inset consisting of three arrows with a half of characteristic tilt angle θ shows that the ice lamellar tip growth velocity $\vec{V}_{tip}$ is the combined result of growth velocities in both basal and edge plane directions. PVA impurity is speculated to affect basal plane growth kinetics. Tilting of ice lamellar tip is influenced by both PVA impurity and $\vec{V}_{pulling}$.

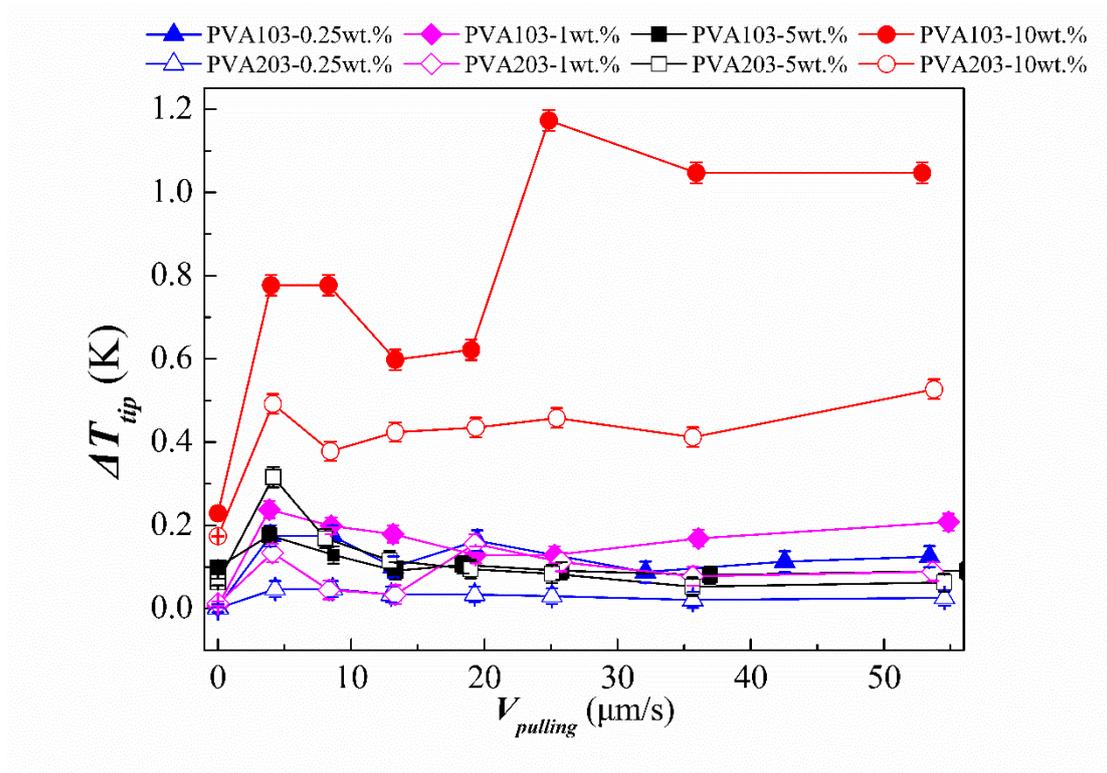

**Figure 10** Tip undercooling of all PVA samples against increasing pulling velocities under steady state.